\documentclass[onecolumn,showpacs,preprintnumbers,amsmath,amssymb]{revtex4}

\usepackage{graphicx}
\usepackage{dcolumn}
\usepackage{amsbsy}

\usepackage{latexsym}
\usepackage{amssymb}
\usepackage{amsmath}

\begin{document}
\baselineskip 12pt

\title{Weak antiferromagnetic ordering and pure magnetic reflections induced by
Dzyaloshinskii--Moriya interaction in MnSi-type
crystals}

\author{Vladimir E. Dmitrienko\footnote{email: dmitrien@crys.ras.ru},
Viacheslav A. Chizhikov} \affiliation{A.V.~Shubnikov Institute of
Crystallography, 119333 Moscow, Russia}


\begin{abstract}
Symmetry analysis of the Dzyaloshinskii-Moriya
(DM) interaction in MnSi-type cubic crystals
demonstrates that the magnetic moments are tilted
periodically, producing a weak antiferromagnetic pattern, when the helix is unwound by
magnetic field. The tilt angles of four Mn
sublattices are determined by a component of the
DM vector perpendicular to that one responsible for
helical spiraling; both components have been evaluated
using a simple model. It is shown that the
tilting should induce pure magnetic reflections
$00\ell,\ell=2n+1$ in neutron or x-ray magnetic
scattering, and the structure factors of these
``forbidden'' reflections are calculated for
arbitrary field orientations.
\end{abstract}
\maketitle

The crystal magnetism is closely related to crystal structure but the details of this relation are not obvious even for simple cases.  
Meanwhile a more clear understanding of microscopic interactions behind numerous types of magnetic ordering would be crucial for magnetic crystal engeneering and applications. Interplay between crystal structure and magnetism
becomes even more intriguing in noncentrosymmetric
crystals where two other players, chirality and
ferroelectricity, can be also very important.
A bright example is the Dzyaloshinskii--Moriya (DM) interaction \cite{Dzyaloshinskii57,Dzyaloshinskii58,Moriya60a,Moriya60b}: being simple in form, it can result in very different magnetic phenomena: weak ferromagnetism in rhombohedral oxides \cite{Turov2010}, multiferroics and exotic magnetics in rare-earth compounds \cite{Sergienko2006,Cheong2007}, helical and Skyrmion structures in MnSi-type crystals \cite{Rossler2006,Grigoriev2006a,Tewari2006}, {\em etc}. It is well understood that this interaction is related with asymmetry of atomic arrangements. However, in spite of numerous theoretical and experimental studies, we have only qualitative understanding of these phenomena rather than quantitative descriptions from the first principles. To move further it would be important to develope new quantitative methods for experimenal studies revealing the physics of underlying magnetic interactions. Recently we suggested \cite{Dmitrienko2010} and realized \cite{Collins2011} a method for experimental determination of the sign of the DM vector in the weak ferromagnetics, and compared the results for FeBO$_3$ with {\em ab initio} simulations. 

In this paper we show that the picture of magnetic ordering in non-centrosymmetric crystals of the MnSi type is more rich than it was thought before. Even in a strong magnetic field when the helical order disappears, there is still a periodic tilting of magnetic moments induced by some component of the DM vector. We estimate the value of this component from a simple model and discuss how it can be measured using x-ray or neutron diffraction.

To find a magnetic pattern formed by four spins $\mathbf{s}^t$ in a unit cell, the classical-spin model, frequently accepted for the magnetic energy $E$ with the DM 
term \cite{Turov2010,Sergienko2006,Cheong2007,Hopkinson2009}, will be used in the following form
\begin{equation}
E = \sum_{t=1}^4\{\frac{1}{2}\sum_{j(t)}\left(-J\mathbf{s}^t \cdot \mathbf{s}^j +
\mathbf{D}^{tj} \cdot [ \mathbf{s}^t \times
\mathbf{s}^j ] \right) -\mu_s \mathbf{H} \cdot
\mathbf{s}^t \}, \label{eq:energy1} 
\end{equation}
where $J>0$ is the
exchange interaction constant, $\mathbf{D}^{tj}$
are the DM vectors of the bonds connecting each of four spins $\mathbf{s}^t$ with its six neighbors, and
$\mu_s$ is an effective magnetic moment of Mn atoms.

For zero magnetic field the ground state of the system is a spin helix which can be reorientated rather easily and becomes conical in not very strong fields. When the helix is unwound by a strong
enough external magnetic field ($|\mathbf{H}|\ge H_{c2}$ \cite{Grigoriev2006a,Grigoriev2006}), the moments are aligned
periodically but, contrary to conventional
ferromagnets without DM interaction, the moment
directions $\mathbf{s}^{t}$ may be slightly
different for four different types of equivalent
atomic positions in the MnSi unit cell
\cite{Chzhikov2011}. For simplicity the temperature dependence of the moments is not taken into account, all the moments are
normalized so that $|\mathbf{s}^t|=1$. The upper index
$t = 1,2,3,4$ enumerates four atomic positions in
the unit cell: $\mathbf{r}^1=(x,x,x)$,
$\mathbf{r}^2=(-x, \frac{1}{2}+x,
\frac{1}{2}-x)$, $\mathbf{r}^3=(\frac{1}{2}-x,
-x, \frac{1}{2}+x)$,
$\mathbf{r}^4=(\frac{1}{2}+x, \frac{1}{2}-x,
-x)$, correspondingly.

Supposing that the DM coefficient is smaller than exchange one ($|\mathbf{D}^{tj}| \ll J$), we can assume that $\mathbf{s}^t$ are tilted from $\mathbf{s}=\mathbf{H}/|\mathbf{H}|$ by small angles $\boldsymbol{\varphi}^t$, $\boldsymbol{\varphi}^t\perp\mathbf{s}$, and express the energy of the unit cell of the crystal as
\begin{equation}
\label{eq:Ecell}
E = \sum_{\{ij\}} ( \frac{J}{2} \boldsymbol{\varphi}^{ij2} + \mathbf{D}^{(ij)} \cdot \boldsymbol{\varphi}^{ij}) + \frac{\mu_s H}{2} \sum_{i=1}^{4} \boldsymbol{\varphi}^{i2},
\end{equation}
where the first summation $\{ij\}$ is taken over twelve differently directed bonds with DM vectors $\mathbf{D}^{(ij)}$ listed in Table \ref{tableD} and $\boldsymbol{\varphi}^{ij} = \boldsymbol{\varphi}^j - \boldsymbol{\varphi}^i$.

\begin{table}[h]
\caption{\label{tableD}Twelve bonds between neighboring Mn atoms and corresponding Dzyaloshinskii--Moriya vectors. The bond $ij$ is directed from the atom of type $i$ to the atom of type $j$. An arbitrary DM vector is assumed for bond $12$ and all other bonds (and their DM vectors) are obtained by symmetry transformations of the cubic point group $23$. In MnSi crystal $x=0.138$.}
\begin{tabular}{rcc}
\hline
\hline
$ij$ & $\mathbf{r}^{ij}$ & $\mathbf{D}^{(ij)}$ \\
\hline
$12$ & $(-2x, \frac{1}{2}, \frac{1}{2}-2x)$ & $(D_x, D_y, D_z)$ \\
$13$ & $(\frac{1}{2}-2x, -2x, \frac{1}{2})$ & $(D_z, D_x, D_y)$ \\
$14$ & $(\frac{1}{2}, \frac{1}{2}-2x, -2x)$ & $(D_y, D_z, D_x)$ \\
$21$ & $(2x, \frac{1}{2}, -\frac{1}{2}+2x)$ & $(-D_x, D_y, -D_z)$ \\
$23$ & $(-\frac{1}{2}, \frac{1}{2}-2x, 2x)$ & $(-D_y, D_z, -D_x)$ \\
$24$ & $(-\frac{1}{2}+2x, -2x, -\frac{1}{2})$ & $(-D_z, D_x, -D_y)$ \\
$31$ & $(-\frac{1}{2}+2x, 2x, \frac{1}{2})$ & $(-D_z, -D_x, D_y)$ \\
$32$ & $(-\frac{1}{2}, -\frac{1}{2}+2x, -2x)$ & $(-D_y, -D_z, D_x)$ \\
$34$ & $(2x, -\frac{1}{2}, \frac{1}{2}-2x)$ & $(-D_x, -D_y, D_z)$ \\
$41$ & $(\frac{1}{2}, -\frac{1}{2}+2x, 2x)$ & $(D_y, -D_z, -D_x)$ \\
$42$ & $(\frac{1}{2}-2x, 2x, -\frac{1}{2})$ & $(D_z, -D_x, -D_y)$ \\
$43$ & $(-2x, -\frac{1}{2}, -\frac{1}{2}+2x)$ & $(D_x, -D_y, -D_z)$ \\
\hline
\hline
\end{tabular}
\end{table}

The minimization of (\ref{eq:Ecell}) gives
\begin{equation}
\label{eq:phi}
\boldsymbol{\varphi}^t = \frac{D_x + D_z}{4J + \mu_s H/2} \left(\boldsymbol{\psi}^t-\mathbf{s}(\boldsymbol{\psi}^t\cdot\mathbf{s})\right),
\end{equation}
where vectors
$\boldsymbol{\psi}^t$ are directed along
threefold axes passing through $\mathbf{r}^t$ and
interrelated by the symmetry transformations of
the space group $P2_13$:
\begin{subequations}     
\label{eq:psi}
\begin{eqnarray}
\boldsymbol{\psi}^1=(1,1,1), \\
\boldsymbol{\psi}^2=(-1,1,-1),\\
\boldsymbol{\psi}^3=(-1,-1,1),\\
\boldsymbol{\psi}^4=(1,-1,-1).
\end{eqnarray}
\end{subequations}

Thus, in the first approximation, the four spins are:
\begin{equation} \label{eq:st2}
\mathbf{s}^t=\mathbf{s}+\delta [\boldsymbol{\psi}^t\times\mathbf{s}] , \phantom{x} \delta=\frac{D_x+D_z}{4J},
\end{equation}
where the magnetic term is neglected in comparison with exchange, $\mu_s H \ll 8J$.
From this equation one can see that four components
$[\boldsymbol{\psi}^t\times\mathbf{s}]$ are
always aligned antiferromagnetically, so that
$\sum_{t=1}^4[\boldsymbol{\psi}^t\times\mathbf{s}]=0$ for any field direction.
They give no contribution to the macroscopic
magnetization, but they can be measured by
diffraction. In particular, they induce pure
magnetic reflections in x-ray or neutron magnetic
scattering.

Indeed, let us consider neutron scattering
(elastic and coherent). It is determined by
the structure factor which is a sum of nuclear and magnetic structure factors
$F(\mathbf{h})=F_{nuc}(\mathbf{h})+F_{mag}(\mathbf{h})$ where $\mathbf{h}=2\pi (h,k,\ell)$ is the
scattering vector (in our case it is the
reciprocal lattice vector with the Miller indices
$hk\ell$).
For simplicity it is supposed that magnetic moments are centered on Mn atoms and possible itinerant nature of MnSi magnetism is ignored. In this case $F_{mag}(\mathbf{h})$ can be written as a sum over all local moments with
corresponding phase multipliers
\begin{equation} \label{eq:Fmag}
F_{mag}(\mathbf{h})=b_m\mathbf{P}\cdot
\sum_{t=1}^4\left(\mathbf{s}^t-\mathbf{h}(\mathbf{h}\cdot\mathbf{s}^t)/\mathbf{h}^2\right)
\exp(i\mathbf{h}\cdot\mathbf{r}^t)
\end{equation}
where $\mathbf{P}$ is the direction
of neutron spin, $b_m$ is the amplitude of neutron magnetic scattering which is proportional to $\mu_s$, to magnetic formfactor $f_m(\mathbf{h})$ and to Debye--Waller factor $f_{DW}$. Taking into account that in the
magnetic field $\mathbf{P}$ is either
parallel or antiparallel to the field direction
so that
$P=\mathbf{P}\cdot\mathbf{s}=\pm 1$
and $\mathbf{P}\cdot
[\boldsymbol{\psi}^t\times\mathbf{s}]=0$ we
obtain
\begin{subequations}     
\begin{eqnarray}
F_{mag}(\mathbf{h})=F_{FM}+F_{AFM} \nonumber\\
=b_m P\left(1-(\mathbf{h}\cdot\mathbf{s})^2/\mathbf{h}^2\right)
\sum_{t=1}^4 \exp(i\mathbf{h}\cdot\mathbf{r}^t) \\
+\delta b_m P\frac{\mathbf{h}\cdot\mathbf{s}}{\mathbf{h}^2}
[\mathbf{h}\times\mathbf{s}]\cdot\sum_{t=1}^4
\boldsymbol{\psi}^t\exp(i\mathbf{h}\cdot\mathbf{r}^t) .
\end{eqnarray} \label{eq:FM-AFM}
\end{subequations}
We see that the antiferromagnetic contribution to the structure factor,
Eq. (\ref{eq:FM-AFM}b), induced by the DM
interaction, vanishes if the direction of
magnetic field is either parallel or
perpendicular to the scattering vector whereas
the ferromagnetic contribution, Eq.
(\ref{eq:FM-AFM}a), as it is well known, vanishes
if the field is parallel to $\mathbf{h}$. However, owing to $2_1$ screw axes, both ferromagnetic and nuclear contributions to the structure factor should  disappear for special type of reflections,  $00\ell,\ell=2n+1$ (forbidden reflections) because for them $\sum_{t=1}^4 \exp(i\mathbf{h}\cdot\mathbf{r}^t)=0$. On the contrary, the antiferromagnetic contribution (\ref{eq:FM-AFM}b), proportional to $\sum_{t=1}^4 \boldsymbol{\psi}^t\exp(i\mathbf{h}\cdot\mathbf{r}^t)$, does not vanish for these reflections and the structure factor is given by 
\begin{eqnarray} \label{eq:FAFM}
F(00\ell,\ell=2n+1)=F_{AFM}(00\ell,\ell=2n+1) \nonumber\\
=4b_m P\delta s_z(-s_y \cos 2\pi\ell x + i s_x \sin 2\pi\ell x).
\end{eqnarray}

The intensity of these reflections, which is proportional to $|F(\mathbf{h})|^2$, can be strongly changed by rotation of the field. It is easy to prove that for allowed reflections of this type, $00\ell,\ell=2n$, the antiferromagnetic part gives no contribution because vector $\sum_{t=1}^4 \boldsymbol{\psi}^t\exp(i\mathbf{h}\cdot\mathbf{r}^t)$ is parallel to $\mathbf{h}$. For oblique directions of $\mathbf{h}$ and $\mathbf{s}$ all the contributions to the structure factor are nonzero and interference between them allows us to measure the absolute value and the sign of $\delta$.

There is another phenomenon which can contribute to the forbidden reflections in MnSi, the anisotropy of atomic susceptibility tensors \cite{Gukasov2002,Cao2009}, previously observed in Nd$_{3-x}$S$_{4}$, Sm$_{3}$Te$_{4}$, and $R_2$Ti$_2$O$_7$ crystals ($R$=Ho, Tb, Er, Yb). The magnetic susceptibility of each atom can be considered as a second-rank tensor $\chi_{jk}$ obeying the point symmetry of the atomic site. The magnetic field 
induces the atomic magnetic moment $M_j=\chi_{jk}H_k$ even in the paramagnetic phase. For rare earths the values of the induced moments can be of about $\mu_B$ in fields of about one Tesla \cite{Gukasov2002,Cao2009}; for transition metals they are never measured. 

It is important that in complex structures the induced moments have different orientations for different atomic sites and can contribute to the neutron scattering amplitude of forbidden reflections. In MnSi, where Mn atoms are at sites with threefold symmetry axes of four different orientations $\boldsymbol{\psi}^t$, the site susceptibility tensor can be written as
\begin{equation} \label{eq:ksi}
\chi_{jk}^t=\chi_{0}\delta_{jk}+\chi_{a}(\psi_j^t\psi_k^t-\delta_{jk})
\end{equation}
where $\delta_{jk}$ is the unit tensor, $\chi_{0}$ is the isotropic susceptibility whereas $\chi_{a}$ describes the anisotropy. The induced magnetic moment is $(\chi_{0}-\chi_{a})\mathbf{H} +\chi_{a}\boldsymbol{\psi}^t(\boldsymbol{\psi}^t\cdot\mathbf{H})$ and after simple calculations we obtain the following contribution to the structure factor
\begin{eqnarray} \label{eq:FAFMchi}
F_{\chi AFM}(00\ell,\ell=2n+1) \nonumber\\
=4b_m P s_z(s_y \cos 2\pi\ell x + i s_x \sin 2\pi\ell x).
\end{eqnarray}
where $b_m$ is now proportional to $\chi_{a}H$ instead of $\mu_s$. Thus, contrary to $F_{AFM}$, the absolute value of $F_{\chi AFM}$ is growing proportional to $H$, and this allows us to distinguish these two contributions to the reflection intensity.

X-ray scattering is also sensitive to spin distribution \cite{Blume85,Lovesey05,Lovesey96} and can be applied instead of neutrons to this type of studies. It has some important advantages, for instance the polarizations of incident and diffracted beams, $\mathbf{e}$ and $\mathbf{e'}$, are changeable regardless the direction of the magnetic field, and, in addition, interference with resonant scattering \cite{Dmitrienko05} can be quite helpful. In our case of forbidden reflections from MnSi-type crystals, two contributions to the x-ray structure factor are most relevant, nonresonant magnetic scattering and resonant dipole-dipole scattering. The latter is expected to be stronger than the former near the $K$-edge but practically disappears far from the edge. Thus, changing the x-ray energy $\hbar\omega$, we can select the best conditions for interference between the two contributions.  

The dipole-dipole tensor of atomic scattering factor looks exactly like Eq. (\ref{eq:ksi}): $f_{jk}^t=f_{0}\delta_{jk}+f_{a}(\psi_j^t\psi_k^t-\delta_{jk})$. For the ``forbidden'' reflections $00\ell,\ell=2n+1$ the resonant dipole-dipole scattering changes polarization from $\boldsymbol{\sigma}$ to $\boldsymbol{\pi'}$ or from $\boldsymbol{\pi}$ to $\boldsymbol{\sigma}$ \cite{Dmitrienko1984} and we consider only these diffraction channels 
\begin{subequations}     
\begin{eqnarray}
F_{\sigma\pi'}(00\ell,\ell=2n+1)=F_{dd}+F_{AFM} \nonumber\\
=f_a\sum_{t=1}^4(\boldsymbol{\pi'}\cdot\boldsymbol{\psi}^t)(\boldsymbol{\psi}^t\cdot\boldsymbol{\sigma}) \exp(i\mathbf{h}\cdot\mathbf{r}^t) \\
+i\delta\mu_{s}f_m(\mathbf{h}) \frac{\hbar\omega}{mc^2}[\mathbf{s}\times\mathbf{B}]\cdot\sum_{t=1}^4
\boldsymbol{\psi}^t\exp(i\mathbf{h}\cdot\mathbf{r}^t),
\end{eqnarray} \label{eq:FM-AFMx}
\end{subequations}
where $\mathbf{B}=[\boldsymbol{\pi'}\times\boldsymbol{\sigma}]+ [\mathbf{k'}\times\boldsymbol{\pi'}](\mathbf{k'}\cdot\boldsymbol{\sigma})- [\mathbf{k}\times\boldsymbol{\sigma}](\mathbf{k}\cdot\boldsymbol{\pi'})- [\mathbf{k'}\times\boldsymbol{\pi'}]\times[\mathbf{k}\times\boldsymbol{\sigma}]=2\mathbf{k}\sin^2\theta$, $\theta$ is the Bragg angle, $\mathbf{k}$ and $\mathbf{k'}$ are the unit wave vectors of incident and diffracted waves. We consider only spin scattering \cite{Blume1988} whereas orbital scattering is neglected. Finaly we obtain
\begin{subequations}     
\begin{eqnarray}
F_{\sigma\pi'}(00\ell,\ell=2n+1)=F_{dd}+F_{AFM} \nonumber\\
=2f_a\sin2\theta(\cos\varphi\cos2\pi\ell x-i\sin\varphi\sin2\pi\ell x) \\
-2\delta\mu_{s}f_m(\mathbf{h}) \frac{\hbar\omega}{mc^2}\sin^2\theta
[(s_x\sin\theta+s_z\cos\theta\cos\varphi)\sin2\pi\ell x  \nonumber\\
+ i(s_y\sin\theta+s_z\cos\theta\sin\varphi)\cos2\pi\ell x], 
\end{eqnarray} \label{eq:FM-AFMx2}
\end{subequations}
where $\varphi$ is the azimuthal angle of $\mathbf{k}$, starting from the $x$-axis. We see that both $F_{dd}$ and $F_{AFM}$ have pronounced azimuthal dependence and $F_{AFM}$ can be changed by changing $\mathbf{s}$. There are quite reliable codes for calculation of energy dependence of $f_a$, for instance \cite{Joly01,Gougoussis09}, so that one can, in principle, obtain both the value and the phase of $\delta$ from interference between $F_{dd}$ and $F_{AFM}$. If magnetic moments have a more complicated distribution, as it is expected for itinerant magnetics, $F_{AFM}$ and $F_{\chi AFM}$ could be quite different from those ones given by Eqs. (\ref{eq:FM-AFMx2}b) and (\ref{eq:FAFMchi}).

We consider in detail only forbidden reflections, but their intensity is expected to be rather small. Perhaps it would be more appropriate to refine the magnetic structure using many nonforbidden reflections, as it has been done in \cite{Gukasov2002,Cao2009}, because from this refinement one can deduce not only the absolute values but also the signs of $D_x+D_z$, $\chi_a$, and $f_a$. 
Besides, the forbidden reflections can be excited owing to magnitostriction: for oblique direction of the magnetic field the space group reduces from cubic to triclinic, the atoms change slightly their positions, and previously ``forbidden'' reflections become allowed. 

In the absence of {\it ab initio} calculations, we can estimate a possible value of the DM vector from a simple expression suggested many years ago by Keffer \cite{Keffer}. Starting from the Moriya theory \cite{Moriya60b} of superexchange coupling  via a near-neighbor intermediate nonmagnetic atom, he had shown that the DM vector could be written as
\begin{equation} \label{eq:keffer}
\mathbf{D}=D[\mathbf{r}^{1i}\times\mathbf{r}^{2i}],
\end{equation}
where $D$ is an unknown coefficient. As above, this DM vector corresponds to the Mn-Mn bond directed from $\mathbf{r}^{1}$ to $\mathbf{r}^{2}$ and vectors $\mathbf{r}^{1i}$ and $\mathbf{r}^{2i}$ are directed from $\mathbf{r}^{1}$ and $\mathbf{r}^{2}$ to the position of the intermediate atom. In the MnSi structure, there are three intermediate Si atoms per each Mn-Mn bond but two of them form an almost perfect parallelogram with two Mn atoms. Therefore their contributions are mutually compensated and we should consider only the contribution of one asymmetric Si atom. For this bond it is positioned at $(-x_{Si},\frac{1}{2}+x_{Si},\frac{1}{2}-x_{Si})$, and Eq. (\ref{eq:keffer}) gives after simple calculations
\begin{equation} \label{eq:keffer2}
\mathbf{D}=\frac{1}{2}D|x-x_{Si}|\left(2-4|x|,1,4|x|-1\right),
\end{equation}
This equation is written in the form valid for the MnSi structures of the opposite chirality which are related by simultaneous inverting the signs of $x$ and $x_{Si}$.

We can find the value of $D$ from the magnetic helix period which is about 40 unit cells (180 \AA); thus, the helix wave vector $k\approx 2\pi/40\approx 0.157$. It was shown \cite{Chzhikov2011} that $k$ is related with the DM vector as $k=2|D_x-2D_y-D_z|/(3J)$; thus, according to Eq. (\ref{eq:keffer2}), $k=|D(x-x_{Si})(1-8|x|)|/(3J)$. Accepting $x=0.138$ we have $|D||x-x_{Si}|/J=4.53$ and $|\delta|=|D_x+D_z|/(4J)=|D||x-x_{Si}|/(8J)\approx 0.57$. We see that parameter $\delta$, which determines the tilt angles, is not very small even in MnSi where the helix period is quite long (for comparison, in MnGe it can be about six unit cells \cite{Kanazawa2011}). Of course, so large value of $\delta$ seems to be a result of our simple model, Eq. (\ref{eq:keffer}), forcing the DM vector to be perpendicular to the bond and ignoring the itinerant nature of magnetism in MnSi. However, the DM vector is perpendicular or almost perpendicular to the bond in more sophisticated calculations \cite{Sergienko2006,Cheong2007,Moskvin1977,Shekhtman93,Mazurenko05,Katsnelson10}. Moreover, according to Eq. (\ref{eq:keffer2}), the third component of the DM vector, $D_x+D_y-D_z$ is unnaturally large, $|D_x+D_y-D_z|/J=2|D||x-x_{Si}|(1-2|x|)/J=6.56$. To clarify all those points it would be very important to have {\it ab initio} calculations of the DM vector for MnSi-type structures.

In conclusion we have demonstrated that the picture of magnetic ordering in MnSi-type crystals is nontrivial even after unwinding of the helix by external field. The remaining antiferromagnetic ordering can be measured by diffraction methods providing information about the component of the Dzyaloshinskii--Moriya vector never measured before. Our na\"{i}ve estimation shows that this component can be quite large and this is a challenge both to experiment and to more reliable calculations.


We are grateful to P.~J. Brown, S.~V. Grigoriev and S.~V.
Maleyev for useful discussions. This work was
supported by two basic research programs of the
Presidium of the Russian Academy of Sciences:
``Thermophysics and Mechanics of Extreme Energy
Actions and the Physics of a Strongly Compressed
Substance'' and ``Resonant X-ray Diffraction and
Topography in Forbidden Reflections''.

\end{document}